\RequirePackage{fix-cm}
\documentclass[smallextended]{svjour3}       
\smartqed  
\usepackage{graphicx}
\begin{document}

\title{Ultrastrong coupling in a scalable design for circuit QED with superconducting flux qubits
}


\author{Mun Dae Kim
}


\institute{Mun Dae Kim \at
Korea Institute for Advanced Study, Seoul 130-722, Korea  \\
              \email{mdkim@kias.re.kr}           
}

\date{Received: date / Accepted: date}

\maketitle

\begin{abstract}
We theoretically study a circuit quantum electrodynamics (QED)
architecture with superconducting flux qubits. The qubit is
coupled to the transmission line resonator by an ac current
originating from the current mode of the resonator. Ultrastrong
coupling can be obtained by varying the capacitance between the
qubit and the resonator. We propose a scalable design where the
two-qubit coupling can be achieved.
\end{abstract}

\section{Introduction}
\label{intro}

An artificial two level system can be coupled with the quantized
electromagnetic field in a superconducting transmission line
resonator, while a natural atom is coupled with cavity. This
circuit quantum electrodynamics (QED) architecture
\cite{Blais,Blais07} is a solid-state analog of cavity QED,
providing a strong coupling strength between the qubit and
resonator owing to the large dipole moment of the artificial
qubit. The circuit QED scheme has been applied to
superconducting qubits among which the flux
qubit \cite{Mooij,Orlando,Kim2} has the advantage of fast gate
operation because the flux qubit does not require low
anharmonicity for long coherence time. There have been many
studies for the circuit QED  with the superconducting flux qubit
\cite{Lind,Oelsner}.
However, the inductive coupling between the flux qubit and the
transmission line resonator of the circuit is too weak to perform
the quantum gate operation.

Recently a galvanic coupling scheme for the circuit QED with the
flux qubits has been proposed to enhance the coupling strength by
sharing the flux qubit loop with the resonator transmission line
\cite{Abd,Niem,Diaz}. On the other hand, for the scheme with ac
current coupling between the flux qubit and the transmission line
resonator \cite{Kim,Steffen,Chow,Tsai}, the qubit and the
resonator are not galvanically coupled with each other, but by a
ac current flowing through the capacitance between the qubit and
the resonator.
In this scheme the three-junctions flux qubit is coupled by ac
current, similarly to the superconducting phase qubit
\cite{Martinis,QIP,Berkley,Sill}. While the states of phase qubit
are defined in terms of the phase degrees of freedom in a
washboard type potential, the present flux qubit uses the
persistent current states as qubit states. The qubit state
preparation and the quantum gate operation are achieved by the ac
current. Indeed, the present flux qubit thus has the advantages of
phase qubit such as fast qubit operation and readout, individual
addressing, and scalability.

This ac current coupling has recently been implemented in
experiments where three-junctions flux qubit is located at the end
of the resonator \cite{Steffen,Chow,Tsai}. In this study, we
introduce a scalable design where qubits are located at the nodes
of the resonator current mode so that many qubits can be
controlled by using a higher harmonic mode of resonator current. We
propose a  qubit scheme which can provide ultrastrong coupling.
The ultrastrong coupling can perform numerous quantum optics
effects in solid state device \cite{Diaz,Ball} and provide a long
coherence time \cite{Nataf} in addition to a fast qubit operation
\cite{Romero}.
In the qubit scheme the capacitance between the qubit and the resonator
can be controlled by varying the width of capacitance line, extended from the qubit loop,
and the distance between the capacitance line and the transmission
line resonator. In the scalable design the qubit-resonator coupling depends on the
number of qubits in the circuit. By varying number of qubits
we calculate the qubit-resonator coupling which shows a maximum of
ultrastrong coupling with reasonable parameter
values. Further we analyze the xy-type interaction  between two qubits
which  can also reach  ultrastrong coupling regime.

\section{ Circuit-QED with superconducting flux qubits}
\label{sec2}

Usually the transmon qubit is coupled with  the voltage mode of
the transmission line resonator through a capacitance
\cite{Blais,Blais07,Majer}. For superconducting flux qubits, there have
been many studies to couple the flux qubit with the current mode
of the transmission line resonator by using mutual inductance
between the qubit loop and the resonator  \cite{Lind,Oelsner} or
by sharing the qubit loop with the resonator \cite{Abd,Niem,Diaz}.
On the other hand the three-junctions flux qubit can also be
coupled with the transmission line resonator through a
capacitance, but in this case it is coupled with the ac current
from the resonator. Ac current flowing across the capacitance
gives rise to the coupling between the qubit and the resonator
\cite{Kim,Steffen,Chow,Tsai}.

\begin{figure}[b]
\hspace{1.5cm}
\includegraphics[width=0.8\textwidth]{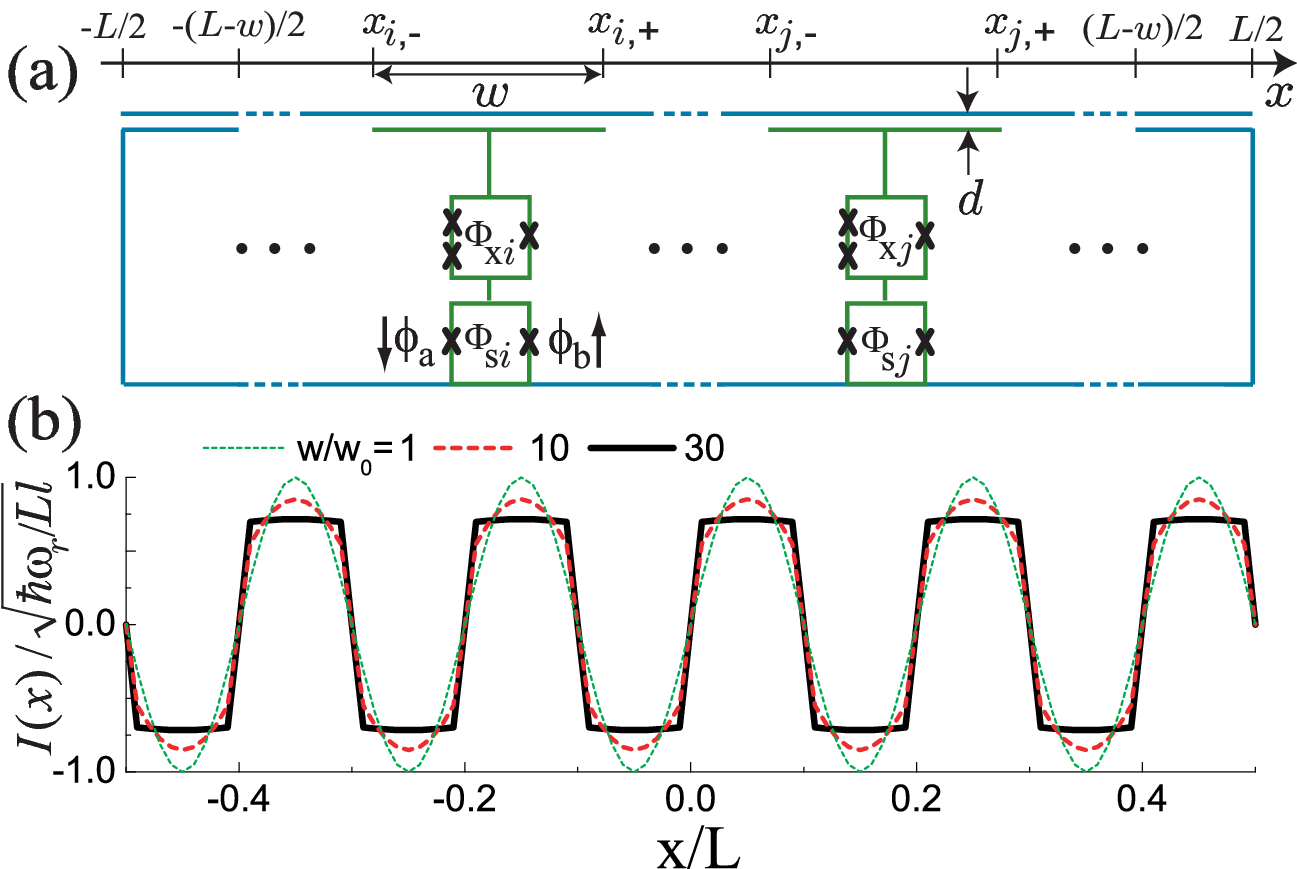}
\vspace{-0cm} \caption{(a) A schematic diagram for a scalable
design of circuit QED with superconducting three-junctions flux
qubits. Here we show, for example, two qubits. The dc-SQUIDs
between qubit and ground plane plays the role of switching the
coupling between qubit and resonator. The capacitance line has the
width of $w$ and the distance between the capacitance line and the
resonator is $d$. (b) Current profiles when there are  nine qubits
in the circuit of (a) for $(d_0/d,w/w_0)=(1,1), (5,10), (15,30)$.
Nine current jumps appear at qubit sites, and grow as the width $w$
of capacitance line increases.} \label{scale}
\end{figure}

We consider a qubit design  shown in Fig. \ref{scale}(a), where the
three-junctions flux qubits are located at the nodes of the current mode
of the resonator. In this scalable design the qubits are coupled with the transmission line
resonator by a capacitance line extended from the qubit loop.
The width $w$ of the capacitance line and the distance $d$ between
the capacitance line and the resonator can be adjusted to
determine the capacitance between the qubit and the resonator.
%
Two capacitors at the ends of the resonator are introduced for the
current mode of the resonator to be periodic in a scalable design.

The microwave passing through the uniform  resonator in the
circuit QED architecture can be described as a one-dimensional
motion by the Lagrangian
\begin{eqnarray}
\label{Lag}
{\cal L}(\theta,{\dot\theta};t)=\int^{\frac{L}{2}}_{-\frac{L}{2}}
\left(\frac{l}{2}({\dot \theta}(x,t))^2-\frac{1}{2c}(\nabla\theta(x,t))^2\right)dx,
\end{eqnarray}
where $l$ and $c$ are the inductance and the capacitance per unit
length of the uniform transmission line resonator, respectively,
and $\theta(x,t)=\int^x_{-L/2}dx'q(x',t)$ with the linear charge
density $q(x)$ is a collective field variable.
In the present design the resonator is not uniform any more. We thus
start with the equation of motion of the field variable in the sector $k$ of the
resonator given by  the Euler-Lagrange equation
\begin{eqnarray}
\label{ELeq}
\frac{1}{c_k}\frac{\partial^2\theta_k(x,t)}{\partial x^2} -l
\frac{\partial^2\theta_k(x,t)}{\partial t^2}=0
\end{eqnarray}
with the capacitance density $c_k$ of sector $k$.

By representing the field variable as a product of  spatial and
temporal parts $\theta_k(x,t)=X_k(x)\phi(t)$, 
we get the equation for the spatial part wavefunction
$(1/c_k)(\partial^2/\partial x^2) X_k(x)+l\omega^2_rX_k(x)=0.$
From this equation we can readily observe that $(1/c_k)\partial
X_k(x)/\partial x$ and $X_k(x)$ are continuous at the boundary
between the sectors, which means the continuity of electric
potential $V_k(x,t)=(1/c_k)\partial\theta(x,t)/\partial
x=(1/c_k)\nabla X_k(x)\phi(t)$ and current
$I(x,t)=\partial\theta(x,t)/\partial t=X(x){\dot \phi}(t)$ at the
boundaries.

\begin{figure}[b]
\hspace{2cm}
\includegraphics[width=0.65\textwidth]{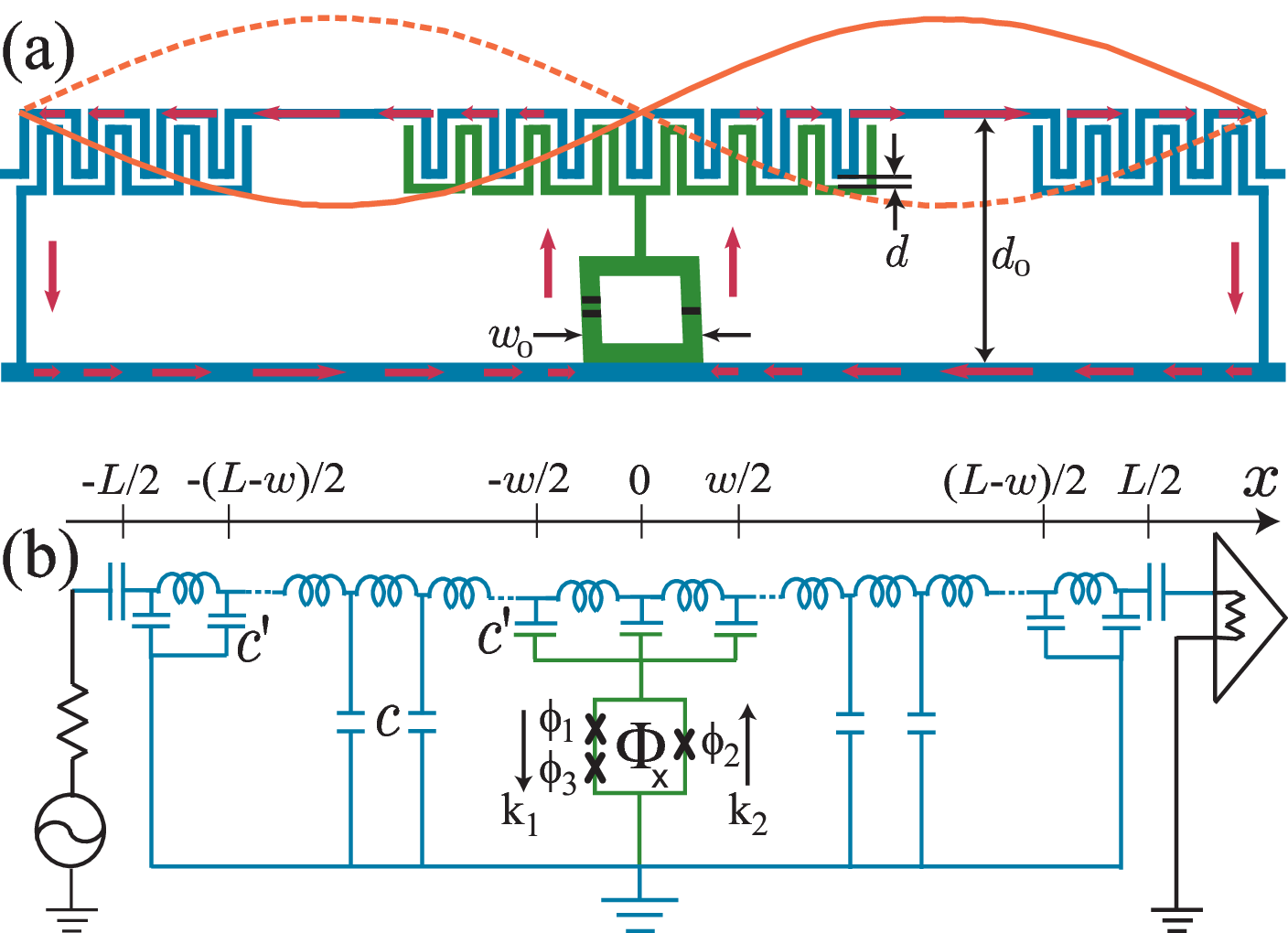}
\vspace{0cm} \caption{ (a) A superconducting three-junctions flux
qubit where $w_0$ is  the width of qubit loop and $d_0$ the
distance between the resonator and the ground plane. The upper
superconducting plane is the transmission line resonator and the
lower is the ground plane. The arrows show the current in the
circuit corresponding to the second harmonic current mode. (b) The
schematic diagram for the circuit in (a), where $c$ and $c'$ are
the capacitance density between the resonator and the ground plane
and between the resonator and the qubit, respectively. The three
Josephson junctions with phase difference $\phi_i$ are located
asymmetrically. } \label{single}
\end{figure}

Let's  consider the case that the number of qubit $N$ is odd.
Then the spatial part $X(x)~(-N-1\leq k \leq N+1)$ of the wavefunction
is written as
\begin{eqnarray}
\label{X}
  X(x)  =  \left\{
\begin{array}{cc}
    A_{ - N - 1}e^{i\frac{j_2\pi}{L}x}  +  B_{ - N - 1}e^{-i\frac{j_2\pi}{L}x}
&      (-\frac{L}{2}<x<-\frac{L}{2}+\frac{w}{2}) \cr
        \vdots & \cr
      A_{k_o}e^{i\frac{j_1\pi}{L}x}  +  B_{k_o}e^{-i\frac{j_1\pi}{L}x} &
        (x_{k_o - 1,+} < x < x_{k_o + 1,-} ) \cr
        \vdots & \cr
    A_{k_e}e^{i\frac{j_2\pi}{L}x} + B_{k_e}e^{-i\frac{j_2\pi}{L}x} &
    (x_{k_e-}<x<x_{k_e+}) \cr
        \vdots & \cr
  A_{N+1}e^{i\frac{j_2\pi}{L}x} + B_{N+1}e^{-i\frac{j_2\pi}{L}x}
&   (\frac{L}{2}-\frac{w}{2}<x<\frac{L}{2}), \cr
\end{array}
\right.
\end{eqnarray}
where
$x_{k\pm}=\frac{k}{N+1}\frac{L}{2}\pm\frac{w}{2}$,
$x_{-N-1,-}=-\frac{L}{2}$, and $x_{N+1,+}=\frac{L}{2}$
with $k_e$ ($k_o$) being even (odd) integer among $k=0,\pm 1,\pm 2,...,\pm (N+1)$.
The capacitance line covers the range $x_{k_e-}<x<x_{k_e+}$ where
a qubit is located. In this region we set $c_k=c'$ and  $j_k=j_2$.
Otherwise, we set $c_k=c$ and  $j_k=j_1$ in the region $x_{k_o - 1,+} < x < x_{k_o + 1,-}$
where there is no qubit [see, for example, Fig. \ref{single}].
From the equation for $X_k(x)$ we obtain
\begin{eqnarray}
\label{omegar}
\frac{1}{\sqrt{lc}}\frac{j_1\pi}{L}=\frac{1}{\sqrt{lc'}}\frac{j_2\pi}{L}
=\omega_r.
\end{eqnarray}

The conditions for continuity of $X(x)$ at, for example,
$x=x_{k_o-1,+}$ and $x=x_{k_e-}$ are given by
\begin{eqnarray}
\label{odd}
&&A_{k_o\!-\!1}e^{i\frac{j_2\pi}{L}x_{k_o\!-\!1,\!+}}\!\!+\!B_{k_o\!-\!1}e^{-i\frac{j_2\pi}{L}x_{k_o\!-\!1,\!+}}
\!\!=\!A_{k_o}e^{i\frac{j_1\pi}{L}x_{k_o\!-\!1,\!+}}\!\!+\!B_{k_o}e^{-i\frac{j_1\pi}{L}x_{k_o\!-\!1,\!+}},\\
\label{even}
&&A_{k_e-1}e^{i\frac{j_1\pi}{L}x_{k_e-}}\!\!+\!B_{k_e-1}e^{-i\frac{j_1\pi}{L}x_{k_e-}}
\!=\!A_{k_e}e^{i\frac{j_2\pi}{L}x_{k_e-}}\!\!+\!B_{k_e}e^{-i\frac{j_2\pi}{L}x_{k_e-}}
\end{eqnarray}
with $k=0,\pm 1,\pm 2,...,\pm N, N+1$, respectively,
and the boundary conditions at both ends of resonator are
\begin{eqnarray}
\label{lend}
&&A_{-N-1}e^{-i\frac{j_2\pi}{2}}+B_{-N-1}e^{i\frac{j_2\pi}{2}}=0,\\
\label{rend}
&&A_{N+1}e^{i\frac{j_2\pi}{2}}+B_{N+1}e^{-i\frac{j_2\pi}{2}}=0.
\end{eqnarray}
From the condition for continuity of $(1/c_k)dX_k(x)/dx$
similar equations are also obtained.

In order to categorize the boundary conditions into even parity and odd parity parts
we set $k_e=-k_o+1$ and use the relation $x_{-k_{e,o},\pm}=-x_{k_{e,o},\mp}$ to
represent, for example, Eq. (\ref{odd}) in terms of $k_e$
and Eq. (\ref{even}) in terms of $k_o$.
Then, all boundary conditions  can be transformed into two sets
with the variables of either (i) $A_{-k}+B_{k}$ or (ii) $A_{-k}-B_{k}$.
Each set can be treated as an independent eigenvalue problem.
If the determinant of the matrix corresponding to set (i) is non-zero while
that for (ii) is zero, we have an odd parity solution such as $A_{-k}=-B_k$.
Around the central qubit site  the solution becomes $X_0(x)\sim A_0 \sin\frac{j_2\pi}{L}x$.
On the contrary, if the determinant for (i) is zero while that for (ii) is non-zero,
we have an even parity solution such as $A_{-k}=B_k$
and thus  $X_0(x)\sim A_0 \cos\frac{j_2\pi}{L}x$.
For the case that the number of qubits $N$ is even,
a similar analysis can also be performed.

The Lagrangian in Eq. (\ref{Lag}) of the resonator modes  can be written as
${\cal L}(\phi,{\dot \phi})= L\left(\frac{l}{2} \mu {\dot
\phi}^2-\frac{1}{2c} \kappa\phi^2\right)$ with dimensionless
constant $\mu=(1/L)\sum_k\int^{L/2}_{-L/2}X^2_k(x)dx$ and
$\kappa=(1/L)\sum_k\int^{L/2}_{-L/2}(c/c_k)(\nabla X_k(x))^2dx$.
If we introduce the representations
\begin{eqnarray}
{\dot \phi}(t)=\frac{-i}{\sqrt{2\mu}}\sqrt{\frac{\hbar\omega_r}{Ll}}(a-a^\dagger),\\
\phi(t)=\frac{1}{\sqrt{2\kappa}}\sqrt{\frac{\hbar\omega_rc}{L}}(a+a^\dagger),
\end{eqnarray}
the Hamiltonian of the resonator modes is written in a
diagonalized form $H_r=\hbar\omega_r(a^\dagger a+\frac12)$. The
current $I(x,t)=X(x){\dot \phi}(t)$ is then given by
\begin{eqnarray}
\label{Ixt}
I(x,t)=-i\frac{X(x)}{\sqrt{2\mu}}\sqrt{\frac{\hbar\omega_r}{Ll}}(a-a^\dagger).
\end{eqnarray}

In Fig. \ref{scale}(b) we show the resonator current profiles for 9 qubits,
which is calculated numerically by using  10th harmonic mode
of the resonator current. In the figure we can observe  current jumps at the
nodes of the current profile where qubits are located.
These jumps increase  along with $w/w_0$ and $d_0/d$ with $w_0$ and
$d_0$ denoted in Fig. \ref{single} (a). The qubit-resonator coupling strength
depends on the current jump as will be seen in the following.

For a single qubit case we can solve the problem analytically. In
Fig. \ref{single}(a)  the qubit is located at the node of
the second harmonic mode of the resonator current.
The resonator and the qubit are coupled by an ac current flowing
into the qubit through the capacitance in the region $-w/2<x<w/2$.
The ac current is given by
\begin{eqnarray}
\label{Ib}
I_b(t)=\int^{w/2}_{-w/2}{\dot q}(x,t)dx=I\left(\frac{w}{2},t\right)-I\left(-\frac{w}{2},t\right)
\end{eqnarray}
from the current conservation condition ${\dot q}(x,t)=\partial
I(x,t)/\partial x$ in the resonator. Beyond the region
$-w/2<x<w/2$ ac current flows from the resonator to the ground
plane directly through the capacitors with small capacitance
density $c$ in Fig. \ref{single}(b).
The amplitude of ac current in Eq. (\ref{Ib}) is given by
\begin{eqnarray}
\label{I0}
I_0=\sqrt{\frac{\hbar\omega_r}{Ll}}\delta
~~~{\rm with}
~~~\delta=2\frac{X(\frac{w}{2})}{\sqrt{2\mu}},
\end{eqnarray}
where $\delta$ corresponds to the current jump in Figs. \ref{scale} and \ref{curr}.
Since the amplitude of current
$I(x,t)$ in Eq. (\ref{Ixt}) satisfies the condition,
$(1/L)\sum_k\int^{L/2}_{-L/2}(X_k(x)/\sqrt{2\mu})^2dx=0.5$,
$\delta$ has the maximum value of $\sqrt{2}$ when the current
profile takes the rectangular function form.

In  Fig. \ref{scale}(a) dc-SQUIDs are inserted between the qubit
and the ground plane for switching on/off the qubit-resonator
coupling. ${\rm \Phi_{xk}}$ and ${\rm \Phi_{sk}}$ are the external
and switching flux for $k$-th qubit, respectively. The present
flux qubit is coupled with the resonator through the ac current
flowing across the capacitance between the qubit and the
resonator. Hence, if we switch  off the ac current by piercing a
half-flux quantum into the dc-SQUID loop, the qubit and the
resonator can be decoupled from each other.
If the self inductance
and the Josephson junction energy of the dc-SQUID loop is as small
as those of the flux qubit, we have the boundary condition
$\phi_a+\phi_b-2\pi\Phi_{si}/\Phi_0\approx 0$ for $i$-th dc-SQUID,
where $\phi_a$ and $\phi_b$ are the phase differences across the
Josephson junctions in the dc-SQUID. Then the total Josephson
junction energy is written as
$E_{JJ}=-E_{J, {\rm eff}}(\Phi_{si})\cos[(\phi_a-\phi_b)/2]$ with
$E_{J, {\rm eff}}(\Phi_{si})=2E_J\cos[(\phi_a+\phi_b)/2]=2E_J\cos(\pi\Phi_{si}/\Phi_0)$, and the current
flow can be blocked for the switching flux $\Phi_{si}=\Phi_0/2$
because $E_{J, {\rm eff}}(\Phi_{0}/2)=0$. This can be understood in another
way: the effective Josephson inductance of the dc-SQUID,
$L_J=(\Phi_0/2\pi)^2/E_{J, {\rm eff}}(\Phi_{si})$, becomes infinite  for
$\Phi_{si}=\Phi_0/2$. Therefore, there can be no current flowing
through, and thus the qubit-resonator coupling is switched off.

However, the boundary condition
$\phi_a+\phi_b-2\pi\Phi_{si}/\Phi_0\approx 0$ is just an approximation
neglecting an induced flux. Including the induced flux due to the current flowing
through the dc-SQUID loop, the boundary condition is replaced by
$\phi_a+\phi_b-2\pi f_t=0$, where $f_t=(\Phi_{si}+L'_sI')/\Phi_0$
with the self inductance $L'_s$ and the current $I'$ of dc-SQUID loop \cite{You2}.
For a sufficiently small dc-SQUID loop the self inductance $L'_s$ and
thus the induced flux $\Phi_{\rm ind}=L'_sI'$ are negligible. However,
in general, since the effective Josephson coupling energy becomes
$E_{J, {\rm eff}}(\Phi_{si})=2E_J\cos[(\phi_a+\phi_b)/2]=2E_J\cos[\pi(\Phi_{si}+L'_sI')/\Phi_0]$,
the external flux should be adjusted as $\Phi_{si}=\Phi_0/2-L'_sI'$
in order to switch off the interaction.

%

\section{Ultrastrong qubit-resonator coupling}
\label{sec3}

When the qubit is located at the center of the resonator, the
spatial part $X(x)$ should be an odd function, because the ac
current of Eq. (\ref{Ib}) vanishes if $X(x)$  is an even function
for $-w/2<x<w/2$. In this case the determinant of matrix
corresponding to set (ii) of boundary conditions in section
\ref{sec2} should be zero, resulting in
\begin{eqnarray}
\label{det}
e^{\frac{ij_1\pi}{2}\left(1-\frac{2w}{L}\right)}=
\pm\frac{cj_2\cos\frac{j_2\pi}{2}\frac{w}{L}-ic'j_1\sin\frac{j_2\pi}{2}\frac{w}{L}}
{cj_2\cos\frac{j_2\pi}{2}\frac{w}{L}+ic'j_1\sin\frac{j_2\pi}{2}\frac{w}{L}}.
\end{eqnarray}
The values of $j_1$, $j_2$ and $\omega_r$ are determined from Eqs.
(\ref{omegar}) and (\ref{det}). For  uniform capacitance density
$c'=c$, $j_1$ and $j_2$ are integers, but in general they are
non-integer depending on the ratio $c'/c$.

The spatial part $X_k(x)~(-2\leq k \leq 2)$ can be obtained by solving
the boundary conditions in section \ref{sec2} for $N=1$.
The set (i) of boundary conditions in section \ref{sec2} provides the
relation $A_{-k}=-B_k$, and the coefficients $A_k$ and $B_k$ are
determined  from the set (ii) of boundary conditions as
\begin{eqnarray}
\label{eqs}
B_{2}    &=&  -e^{ij_2\pi}A_2,\\
A_{ 1}    &=&   - \frac{1}{j_1} e^{ \frac{i\pi}{2}
\left( j_2- j_1( 1- \frac{w}{L} ) \right)}
\left( - \frac{cj_2}{c'} \cos \frac{j_2\pi}{2}\frac{w}{L}
+ ij_1 \sin  \frac{j_2\pi}{2} \frac{w}{L}\right) A_2,\\
B_{ 1}    &=&   - \frac{1}{j_1} e^{ \frac{i\pi}{2}
\left( j_2+j_1( 1- \frac{w}{L} ) \right)}
\left(\frac{cj_2}{c'} \cos \frac{j_2\pi}{2}\frac{w}{L}
+ ij_1 \sin  \frac{j_2\pi}{2} \frac{w}{L}\right) A_2,\\
A_{0}    &=&  \frac{c'j_1}{2cj_2\cos\frac{j_2\pi}{2}\frac{w}{L}}
\left(e^{i\frac{j_1\pi}{2}\frac{w}{L}}A_1-e^{-i\frac{j_1\pi}{2}\frac{w}{L}}B_1\right).
\end{eqnarray}

\begin{figure}[b]
\vspace{4cm}
\includegraphics{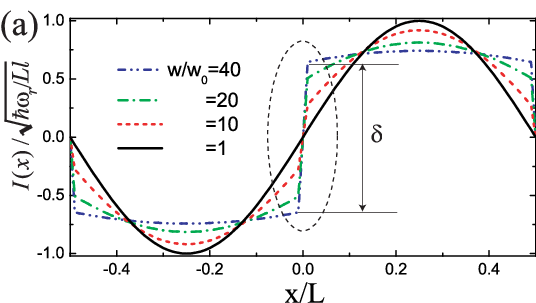}
\vspace{2.5cm}
\includegraphics{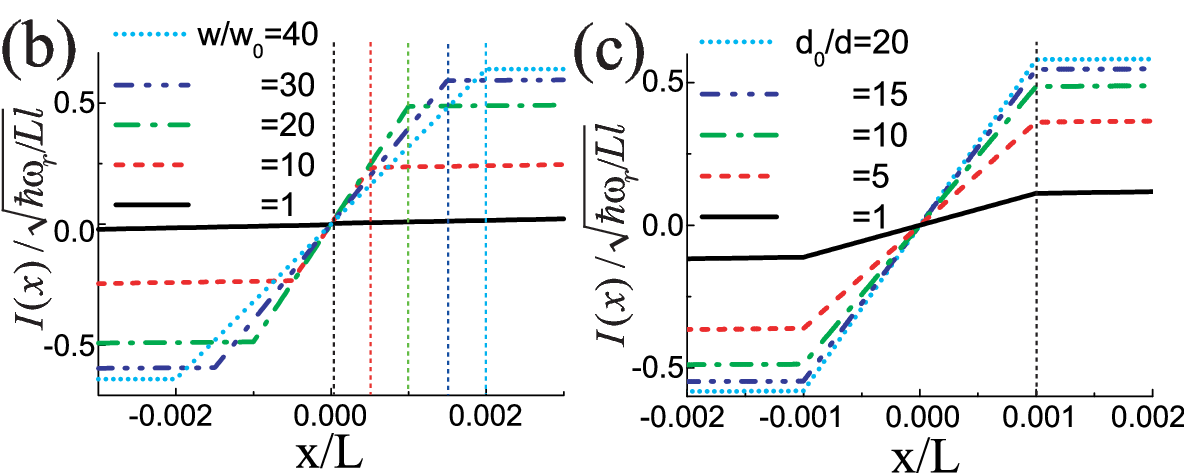}
\vspace{1cm}
\caption{(a) Current profile of single qubit case for various $w/w_0$
with $d_0/d=10$ and  $w_0/L=10^{-4}$.
The amount of jump $\delta$ around $x\approx 0$ increases along with $w/w_0$.
(b) The enlarged current profile around the jump inside the dotted ellipse in (a)
for various $w/w_0$ with $d_0/d=10$ and (c) for various $d_0/d$ with $w/w_0$=20.
Thin dotted lines indicate the boundary position $x=w/2$, the end of the capacitance line.
As $w/w_0$ and $d_0/d$ increase, the jump grows and finally saturates.}
\label{curr}
\end{figure}

In Fig. \ref{curr} (a) we show the current profile
$I(x)=\frac{X(x)}{\sqrt{2\mu}}\sqrt{\frac{\hbar\omega_r}{Ll}}$,
where we set $w_0/L=10^{-4}$ corresponding to $w_0=1\mu$m when $L=10$mm.
In the spatial part wave function $X(x)$ for $N=1$
the remaining coefficient $A_2$ is a common factor in the numerator
and denominator of $I(x)$, and thus is cancelled out.
A finite current jump  develops around the qubit
location at the center of the resonator. Here, for simplicity, we
assume that the capacitance $c'$ between the resonator and the
capacitance line increases linearly along with $w/w_0$ and $d_0/d$
such that $c' = (wd_0/w_0d)c$. When $c' \gg c$,  almost all current flows through the
qubit at the center and the capacitors at the ends of the
resonator, while just a weak current flows directly to ground through the small
capacitance density $c$. As a result, we can observe a sharp change
of current in Fig. \ref{curr}.

Fig. \ref{curr}(b) shows the central part of current profile
closed by a dotted ellipse in Fig. \ref{curr}(a) for various
$w/w_0$ with fixed $d_0/d$, demonstrating a larger jump for larger
$w/w_0$. At the boundary ($x=\pm w/2$), the electric potential
$(1/c_k)\partial X_k(x)/\partial x$ is continuous. Fig.
\ref{curr}(c) shows the currents for various $d_0/d$ with fixed
$w/w_0$, which shows the jump also grows along with $d_0/d$. These
figures show that the current jump $\delta$ grows along with both
$w/w_0$ and $d_0/d$, and  finally saturates at $\sqrt{2}$.

\begin{figure}[b]
\hspace{2cm}
\includegraphics[width=0.7\textwidth]{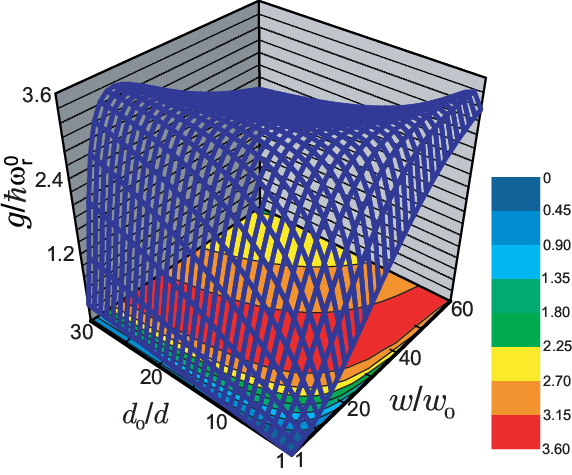}
\vspace{-0cm}
\caption{  Coupling constant $g$ for the single qubit case in the plane of $(d_0/d, w/w_0)$
with the resonator impedance $Z=50\Omega$. $g$ has a maximum value corresponding
to a ultrastrong coupling. At the bottom the contour plot of $g$ is shown. }
\label{g3D}
\end{figure}

The coupling strength between the resonator and the
three-junction flux qubit is given in terms of the amplitude of
the bias-current $I_0$. If the three-junctions flux qubit is
penetrated by a magnetic flux of half flux quantum $\Phi_0/2$,
there are two current states in the qubit loop. The clockwise and
counterclockwise current states correspond to the local minima in
the effective potential of the qubit loop.
Here we consider the usual Josephson junction energies such that $E_{J1}=0.8E_{J2}$ and
$E_{J2}=E_{J3}$ in the three-junction flux qubit of Fig. \ref{single}   \cite{Orlando,Kim2,You}.
In this case the phase difference  $\alpha$ across the small Josephson
junction can be obtained from Eq. (23)
of Ref. \cite{Kim2} with $\lambda=E_{J1}/E_{J2}=0.8$ and $\eta\approx E_L/E_{J2}=50$
with $E_L$ being the characteristic inductive energy, resulting in $\alpha\approx 0.38\pi$.

When  a bias current $I_b$ is applied to the flux qubit where
three Josephson junctions are located asymmetrically in the loop
as shown in Fig. \ref{single}, the Lagrangian can be written as \cite{Kim}
\begin{eqnarray}
\label{Lag}
{\cal L}(\phi_i,\dot{\phi}_i)&=&\sum^3_{i=1}\frac12 C_i\left(\frac{\Phi_0}{2\pi}\right)^2\dot{\phi}^2_i
-U_{\rm eff}(\{\phi_i\}),\\
\label{Ueff}
U_{\rm  eff}(\{\phi_i\})&=&\sum^3_{i=1}E_{Ji}(1-\cos\phi_i)+\frac{\Phi_0I_b}{4\pi}(\phi_1+\phi_3-\phi_2)\nonumber\\
&&+\frac{\Phi^2_0}{2L_s}\left(n+f-\frac{1}{2\pi}\sum^3_{i=1}\phi_i\right)^2,
\end{eqnarray}
where $C_i$ is the capacitance of junction, $L_s$ is the self inductance of the qubit loop, and $f=\Phi_x/\Phi_0$ with an external flux $\Phi_x$ threading the qubit loop and the superconducting unit flux quantum $\Phi_0=h/2e$.

In the present circuit-QED scheme the current flowing into the qubit with width $d$ is given by
$I_b(t)=-iI_0[a(t)-a^\dagger(t)]$.
Then the total Hamiltonian $H_{\rm JC}=H_r+H_q+H_I$ given by the sum of
the Hamiltonian for the resonator mode, for the qubit, and for
the interaction between the resonator mode and the qubit is
written as a Rabi type Hamiltonian \cite{Kim}
\begin{eqnarray}
\label{Hc}
{\cal H}=\hbar\omega_r a^\dagger a
+\frac{\hbar\omega_a}{2}\sigma_z+i g\sigma_x(a-a^\dagger),
\end{eqnarray}
where $\omega_a$ is the qubit frequency and the last term
represents the coupling between the qubit and the current mode in
the resonator  \cite{Kim,Tsai}, which is different from $g\sigma_x(a+a^\dagger)$ in
the transmon case. The coupling strength $g$ is given by a product of  flux variable and bias current
\begin{eqnarray}
\label{g}
g=\frac{\Phi_0}{2\pi}\alpha I_0
\end{eqnarray}
similarly to the superconducting phase qubit
\cite{Berkley,Tinkham}. Here we have the phase difference $\alpha$
because the two phases in different sides of
the qubit cancel each other out. Thus an asymmetrical
layout of Josephson junctions in qubit loop can provide a finite
coupling strength $g$.

Since the amplitude of ac current $I_0$ is given in terms of the current jump
in Eq. (\ref{I0}), the coupling strength $g$ depends on the current jump $\delta$.
The coupling constant  can be rewritten as
\begin{eqnarray}
\label{g2}
\frac{g}{\hbar\omega^0_r}&=&\frac{\alpha\Phi_0}{2\pi}\sqrt{\frac{\omega_r}{\omega^0_r}}
\frac{1}{\sqrt{Ll}}\frac{\delta}{\sqrt{\frac{h}{2\pi}\frac{\pi}{\sqrt{lc}L}}}\nonumber\\
&=&\frac{\alpha}{\pi}\frac{\Phi_0}{\sqrt{hZ}}
\sqrt{\frac{\omega_r}{\omega^0_r}}\frac{\delta}{\sqrt{2}},
\end{eqnarray}
where  $\omega^0_r=\pi/\sqrt{lc}L$ is the frequency of the 1st
harmonic mode of the uniform resonator and
$Z=\sqrt{l/c}$ is the impedance of the resonator \cite{Huang}.
In Fig. \ref{g3D} we show the coupling constant $g$  in the plane
of  ($d_0/d$, $w/w_0$) for $Z=50\Omega$ with $Ll=$2.5nH and $Lc=$1pF \cite{Oelsner,Abd},
and thus $\Phi_0/\sqrt{hZ}\approx 11.37$.

\begin{figure}[b]
\hspace{2cm}
\includegraphics[width=0.55\textwidth]{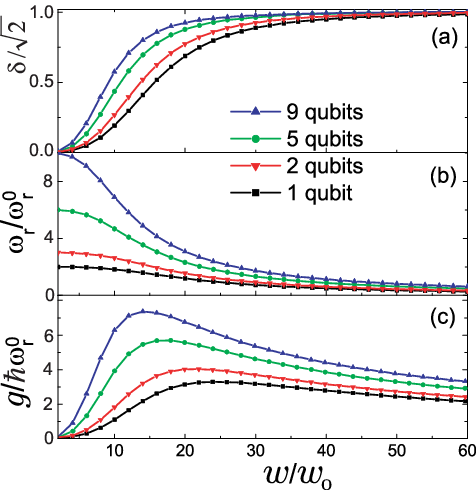}
\vspace{-0cm}
\caption{ (a) The current jump  $\delta$   for $N=$1, 2, 5, 9 qubits
in the circuit of Fig. \ref{scale} as a function of $w/w_0$
along the diagonal line  $w/w_0=2d_0/d$ in the bottom of Fig. \ref{g3D}.
$\delta/\sqrt{2}$ increases along with $w/w_0$ and saturate to 1 finally.
(b) The resonator frequency $w_r$ decreases as $w/w_0$ increases.
(c) The coupling constant $g$ shows a  maximum which
is larger for more number of qubits.}
\label{fig5}
\end{figure}

In the case for more than one qubit in the circuit
we calculate  the coupling strength numerically. 
In Fig. \ref{fig5} the behaviors of  the current jump $\delta$, the
resonator frequency $\omega_r$, and the coupling $g$ along the
diagonal line $w/w_0=2d_0/d$ in the plane of ($w/w_0$, $d_0/d$) in
Fig. \ref{g3D} are shown.
Large capacitance $c'$ enables more
charges to flow across the qubit, resulting in large ac current
and thus large jump in  Fig. \ref{fig5}(a).
However  the frequency of the
resonator mode $\omega_r$ becomes small  for large $c'$. The
increase of $w/w_0$ and $d_0/d$ makes the average capacitance of
the resonator larger, and thus the resonator frequency $\omega_r\sim 1/\sqrt{c'}$
smaller [Fig. \ref{fig5}(b)]. As a result, the coupling $g$ in Eq. (\ref{g2})
demonstrates a maximum because $\omega_r$ decreases while
$\delta$ increases [Fig. \ref{fig5}(c)].
From Figs. \ref{fig5} (b) and (c) it  is shown that
the ultrastrong coupling regime ($g\sim \hbar\omega_r$) is
achievable with $w$ and $1/d$ of just several multiples of $w_0$
and $1/d_0$. If there is one qubit in the circuit, the coupling  shows a  maximum
where the coupling $g\sim 3\hbar\omega^0_r$ reaches  deep strong coupling regime ($g > \hbar\omega_r$) \cite{Ball,Romero,Casanova}.

If there are N qubits in the circuit, we can perform similar
calculations numerically. For N qubits  we need (N+1)th harmonic
mode in the resonator, and thus $\omega_r$ has larger value for
more number of qubits [Fig. \ref{fig5}(b)]. Higher modes have a
shorter wave length, and thus larger amount of ac current flows
from the capacitance line of width $w$ to the qubit. Indeed, the
current jump $\delta$ becomes larger for more number of qubits
[Fig. \ref{fig5}(a)]. As a result, the coupling $g$ in Eq.
(\ref{g2}) increases along with the  number of qubits $N$ [Fig.
\ref{fig5}(c)].
Actually we have calculated the maximum coupling strength
for $N=1,3,5,7,9$ and found that the increasing rate slightly decreases.
In the present scheme the dc-SQUIDs switch on just two qubits
for qubit-qubit interaction, keeping other qubits switched off.
Since only two qubits are coupled in the design, the collected decoherence
rate will be not so much different from that of independent qubit.

There are some disadvantages in our scheme. Actually, we use a higher resonator
mode in the design containing many qubits as shown in Fig. \ref{scale}.
However, it is known that the higher modes of the resonator have low quality factor
which is proportional to 1/N for N-th mode \cite{Goppl}.
In the present study we consider only one mode, namely (N+1)-th mode
for N qubits in one resonator. In order to be more accurate
one should consider the multi mode effect. This effect is manifest,
for example, in the studies of relaxation process of circuit-QED scheme
by the Purcell effect  \cite{Houck} and of multi mode mediated qubit-qubit coupling
\cite{Filipp,McKay}. A fuller treatment of the multi mode effect may be deferred to a future study.

In our scheme the flux qubit is coupled with the transmission line resonator
through a capacitance. The structure of present current-biased flux qubit
coupled with the ac current of resonator is very similar to the usual
current-biased dc-SQUID qubit (phase qubit) \cite{Simmonds}. 
The only difference is the number of
Josephson junctions in the qubit loop, that is, the present qubit has
three Josephson junctions and the phase qubit one or two junctions.
In the present scheme the capacitance is introduced in order to couple
the qubit with external current, and thus the main decoherence may come from the dephasing due to the noise of current flowing through the capacitance. 

The current noise-induced dephasing of the phase qubit is given by
$<\phi^2>\sim(S^*_q ({\rm 1Hz})/C)/\Delta U$ \cite{Martinis2},
where $S^*_q$(1Hz) is the noise spectral density, $C$ the junction capacitance,
and $\Delta U$ the barrier height of tilted potential.
Since the noise spectral density scales as the junction area
which is proportional to the junction capacitance,
$S^*_q ({\rm 1Hz})/C$ is independent of the junction size \cite{Martinis2}.
Thus we can compare the dephasing rates of the present qubit and the phase qubit
by estimating the barrier height of potential, $\Delta U$.
The barrier height of the phase qubit is given by
$\Delta U=(2\sqrt{2} I_0 \Phi_0/3\pi)(1-I/I_0)^{3/2}\approx 2E_J (1-I/I_0)^{3/2}$
with the Josephson coupling energy $E_J=I_0 \Phi_0/2\pi$,
the junction critical current $I_0$,  and the junction bias current $I$
\cite{Martinis2}. Since the junction bias current is typically
driven close to the critical current such as $(1-I/I_0)\sim 10^{-2}$ \cite{Simmonds},
the barrier height becomes very small, $\Delta U\approx 0.002E_J$,
which results in the severe dephasing $<\phi^2>$ in the phase qubit.
However, for our flux qubit we need not the dc-bias
current $I$ and thus  $\Delta U'$ in the double well potential of qubit 
is very large compared to the phase qubit.
For typical flux qubit with smaller junction of $E'_{J1}/E'_{J2}\approx 0.8$
we have $\Delta U'\sim 0.2E'_{J2}$ and the ratio of Josephson coupling energy of
typical flux qubit to that of the phase qubit is $E'_{J2}/E_J\sim 0.1$.
Hence the barrier height of present qubit $\Delta U'/\Delta U \sim 10$,
which results in small dephasing rate compared to the phase qubit.

\section{Two-qubit coupling}

Single qubit gates can be performed by applying an external driving
mode ${\cal H}_D=\epsilon a^\dagger e^{-i\omega_d t}+\epsilon^* a
e^{i\omega_d t}$. The total Hamiltonian is written as
${\cal H}_t={\cal H}_{\rm JC} + {\cal H}_D$, where ${\cal H}_{\rm JC}$
is a Jaynes-Cummings type Hamiltonian
in the rotating wave approximation (RWA) of  ${\cal H}$ in Eq. (\ref{Hc}) \cite{Jaynes}.
By using the transformation  ${\cal D}(\gamma)=e^{\gamma a^\dagger-\gamma^* a }$ with
$\gamma(t)=-(\epsilon/\Delta_r)e^{-i\omega_d t}$ and
$\Delta_r=\omega_r-\omega_d$, we can get the transformed
Hamiltonian ${\tilde {\cal H}}={\cal D}^\dagger {\cal H}_t {\cal D} -i{\cal
D}^\dagger {\dot {\cal D}}$  given by
\begin{eqnarray}
{\tilde {\cal H}}=\Delta_r a^\dagger a+\frac{\Delta_a}{2}\sigma_z
-ig(a^\dagger\sigma_--a\sigma_+)+\frac{\Omega_R}{2}\sigma_y
\end{eqnarray}
with $\Delta_a=\omega_a-\omega_d$.
%
In the dispersive regime $|\Delta| \gg g$ with $\Delta=\omega_a-\omega_r$
the coupling between qubit and resonator can be
eliminated by introducing the transformation ${\cal H}^*={\cal
U}^\dagger{\tilde {\cal H}}{\cal U}$ with ${\cal
U}=e^{-i\frac{g}{\Delta}(a^\dagger\sigma_-+a\sigma_+)}$, resulting in
${\cal H}^*\approx\Delta_r a^\dagger a+\frac{\Delta_a}{2}\sigma_z
+\chi(a^\dagger a +\frac12)\sigma_z+\frac{\Omega_R}{2}\sigma_y$
with the ac-Stark shift $\chi=g^2/\Delta$.

The universal gate in quantum computing requires a two-qubit gate
in addition to the single qubit operation. In the scalable design
of Fig. \ref{scale}(a) the two-qubit Hamiltonian is given by
\begin{eqnarray}
\label{H2q}
H_{\rm 2qubit}  =\omega_r a^\dagger a +   \sum_{j=1,2} \frac{\omega_{aj}}{2}\sigma_{zj}
 - i\sum_{j=1,2}g_j(a^\dagger\sigma_{-j} - a\sigma_{+j})
\end{eqnarray}
in the rotating wave approximation for weak coupling strength
$g_j/\omega_{aj}\ll 1$ \cite{Jaynes}.

The two-qubit Hamiltonian of Eq. (\ref{H2q}) can be represented as
$H_{\rm 2qubit}=H_{\rm cavity}\otimes H_{\rm qubit 1} \otimes
H_{\rm qubit 2}$ and we introduce a transformation matrix
\begin{eqnarray}
\label{calU2} {\cal
U}_2=e^{-i\frac{\varphi_1}{\sqrt{2}}(a^\dagger\sigma_{-1}+a\sigma_{+1})
-i\frac{\varphi_2}{\sqrt{2}}(a^\dagger\sigma_{-2}+a\sigma_{+2})}
\end{eqnarray}
in the same basis. Then the Hamiltonian $H_{\rm 2qubit}$ and the
transformation matrix ${\cal U}_2$ can be written in a
block-diagonal form  by slightly changing the order of basis.
For simplicity, we consider nominally identical qubits,
$\omega_{a1}=\omega_{a2}=\omega_a$,  $g_1=g_2=g$, and thus
$\varphi_1=\varphi_2=\varphi$, and then the lowest block involving
the resonator photon number $n=$ 0 and 1 in the Hamiltonian is
represented in the basis $\{|1\downarrow\downarrow\rangle,
|0\uparrow\downarrow\rangle, |0\downarrow\uparrow\rangle \}$. Here
$|\uparrow\rangle$ and $|\downarrow\rangle$ are the qubit states,
and $|0\rangle$ and $|1\rangle$ are the photon number states.

Then we can easily check that if the condition $\tan
2\varphi=2\sqrt{2}g/\Delta$ is satisfied, the transformed
Hamiltonian ${\tilde H}_{\rm 2qubit}={\cal U}^\dagger_2 H_{\rm
2qubit} {\cal U}_2$ becomes block-diagonalized  further,
and describes the xy-type coupling between two states,
$|0\uparrow\downarrow\rangle$ and $|0\downarrow\uparrow\rangle$,
with the coupling constant
\begin{eqnarray}
J=\frac{g}{\sqrt{2}}\tan\varphi.
\end{eqnarray}
$J$ can be explicitly evaluated with above condition to provide
the interaction Hamiltonian
\begin{eqnarray}
\label{H2i} H_{\rm
int}=\pm\frac{g^2}{\sqrt{\left(\frac{\Delta}{2}\right)^2+2g^2}+\frac{|\Delta|}{2}}
(\sigma_{-1}\sigma_{+2}+\sigma_{+1}\sigma_{-2}),
\end{eqnarray}
where the sign is $+$ for $\Delta>0$ and $-$ for $\Delta<0$
because $g>0$.  In the circuit-QED architecture the interaction between the resonator mode and the qubit  can be discussed separately in the on-resonance regime
($|\Delta|=|\omega_{a}-\omega_r|\approx 0)$ and the dispersive regime ($g/|\Delta| \ll 1$) \cite{Wallraff}.
This expression holds for the on-resonance regime as well as for the dispersive regime.
In the on-resonance regime the interaction term becomes
$H_{\rm int}\approx \pm (g/\sqrt{2})(\sigma_{-1}\sigma_{+2}+\sigma_{+1}\sigma_{-2})$.

In real experiments, however, the two flux qubits may not be identical because of
the different Josephson junction energies between two qubits. Then the qubit frequencies
$\omega_{aj}$ with $j=1,2$  as well as the phase difference
$\alpha_j$'s across the small Josephson junction  are different from each other.
Further, the capacitance  density $c'$ around  each qubit may also be different.
As shown in Fig. \ref{curr} the current jump $\delta$ depends on the
capacitance density $c'$.  As a result, the qubit-resonator coupling strength $g_j$
in Eq. (\ref{g2}) are also different from each other.

For this general case we can obtain the two-qubit interacting Hamiltonian
in the dispersive regime.
Since the condition $\tan 2\varphi=2\sqrt{2}g/\Delta$ can be
reduced to $\varphi \approx \sqrt{2}g/\Delta$ in the dispersive
regime $|\Delta_j|=|\omega_{aj}-\omega_r|\gg g_j$, the
transformation in Eq. (\ref{calU2}) is approximately given by
\begin{eqnarray}
\label{U2}
U_2=e^{-i\frac{g_1}{\Delta_1}(a^\dagger\sigma_{-1}+a\sigma_{+1})
-i\frac{g_2}{\Delta_2}(a^\dagger\sigma_{-2}+a\sigma_{+2})}.
\end{eqnarray}
Then we can obtain the transformed Hamiltonian ${\tilde H}_{\rm 2qubit}={U}^\dagger_2 H_{\rm
2qubit} {U}_2$, resulting in
\begin{eqnarray}
\label{Hint}
H_{\rm int}=\frac12\left(\frac{1}{\Delta_1}+\frac{1}{\Delta_2}\right)
g_1g_2(\sigma_{-1}\sigma_{+2}+\sigma_{+1}\sigma_{-2})
\end{eqnarray}
which can also be seen in the Halimtonian for circuit-QED with
superconducting charge qubits \cite{Blais07}.  In the dispersive
limit  we can check that the Hamiltonian in Eq. (\ref{H2i})
can be reduced to that in Eq. (\ref{Hint}) for identical two qubits with
$\Delta_{1}=\Delta_{2}$ and $g_1=g_2$.

\section{Summary}

In summary, we proposed a circuit QED architecture with
superconducting flux qubits. The three-junctions flux qubit is
coupled with the resonator by an ac current flowing through the
capacitance between the qubit and the resonator. We introduced a
scalable design with superconducting flux qubits, where the
capacitance line extended from the qubit loop  takes the role of
leading the oscillating current into the qubit loop. As the
capacitance between qubit and resonator increases, larger current
flows through the qubit while the resonator mode frequency
decreases. As a result, the qubit-resonator coupling shows a
maximum of ultrastrong coupling with reasonable parameter values.
The qubit-resonator coupling increases along with the number of
qubits because a higher harmonic mode should be adopted.
Two-qubit xy-type interaction can also be obtained from a scalable design
for circuit QED architecture.


\begin{acknowledgements}
The author acknowledges the useful discussion with K. Moon.
This work was partly supported by Basic Science Research Program
through the National Research Foundation of Korea (NRF) funded by
the Ministry of Education, Science and Technology (2011-0023467)
and by the IT R$\&$D program of MOTIE/KEIT [10043464(2012)].
\end{acknowledgements}



Conflict of Interest: The author declares that he has no conflict of interest.



\end{document}